\documentclass[twocolumn]{article}

\usepackage{geometry}
\usepackage[dvipsnames]{xcolor}
\usepackage{soul}

\usepackage{subcaption}
\usepackage{booktabs}
\usepackage{adjustbox}
\usepackage{braket}
\usepackage{multirow}
\usepackage[all]{nowidow}
\usepackage{float}
\usepackage{booktabs}
\usepackage[dvipsnames,table]{xcolor} 

\usepackage{amsmath}
\usepackage{amssymb}
\usepackage{bm}
\usepackage{nicefrac}
\usepackage{mathtools}
\usepackage{booktabs}
\usepackage[separate-uncertainty=false]{siunitx}
\usepackage{pgfplots}
\usepackage{tikz}
\pgfplotsset{compat = newest}
\usetikzlibrary{quotes,angles,calc,arrows,patterns,quotes,external}
\usepgfplotslibrary{groupplots}

\usepackage[colorlinks=true, citecolor=blue,urlcolor=blue,linkcolor=blue,filecolor=black]{hyperref}
\usepackage[capitalise,nameinlink,noabbrev]{cleveref} 

\usepackage{authblk}

\usepackage{algorithm}
\usepackage{algpseudocode}

\usepackage[T1]{fontenc}

\title{Highly-Indistinguishable Single-Photons at 1550 nm from a Two-photon Resonantly Excited Purcell-enhanced Quantum Dot}

\author[1]{Robert Behrends}
\author[1]{Martin v. Helversen}
\author[1]{Pratim K. Saha}
\author[1]{Lucas Rickert}
\author[1]{Koray Kaymazlar}
\author[1]{Mareike Lach}
\author[1]{Nils D. Kewitz}
\author[2,3]{Jochen Kaupp}
\author[2,3]{Yorick Reum}
\author[3,4]{Tobias Huber-Loyola}
\author[2,3]{Sven Höfling}
\author[2,3]{Andreas Pfenning}
\author[5,*]{Tobias Heindel}
\affil[1]{Institute of Physics and Astronomy, Technical Univeristy Berlin, Hardenbergstraße 36, 10623 Berlin, Germany
}
\affil[2]{Würzburg-Dresden Cluster of Excellence ct.qmat, University of Würzburg, Am Hubland, 97074 Würzburg, Germany}
\affil[3]{Technische Physik, Physikalisches Institut, University of Würzburg, Am Hubland, 97074 Würzburg, Germany}
\affil[4]{Institute of Photonics and Quantum Electronics (IPQ) and Center for Integrated Quantum Science and Technology (IQST), Karlsruhe Institute of Technology, Engesserstr. 5, 76131 Karlsruhe, Germany}
\affil[5]{Department for Quantum Technology, Univeristät Münster, Heisenbergstraße 11, 48149 Münster, Germany}
\affil[*]{Corresponding author: tobias.heindel@uni-muenster.de}
\date{\today}

\begin{document}


\twocolumn[
\begin{@twocolumnfalse}
\maketitle
\begin{abstract}
\noindent
In this work we present a cavity-enhanced InAs/$\mathrm{In_{0.53}Al_{0.23}Ga_{0.24}As}$ quantum dot (QD) single-photon source in the telecom C-band with a record-low biexciton emitter decay time of \SI{67.4(2)}{ps} under resonant two-photon excitation (TPE). We observe strong multiphoton suppression associated with $g^{(2)}_\mathrm{X}(0) = 0.006(1)$ and $g^{(2)}_\mathrm{XX}(0) = 0.007(1)$ for the exciton (X) and biexciton (XX) emission, respectively. Due to a asymmetric Purcell enhancement of the XX-X cascade, the two-photon interference (TPI) visibility of XX photons under $\pi$-pulse excitation of $V_{\rm{TPI}} = 90(3)\%$ reaches the theoretical limit and clearly exceeds the $\sim60\%$ expected for standard XX-X cascades without photonic engineering. Furthermore, adding a second timed laser pulse coinciding with XX emission energy, we demonstrate stimulated TPE in the telecom C-Band. The result is an improved TPI visibility of the X photons of $V_{\rm{TPI}}=0.69(3)$ compared to TPE with $V_{\rm{TPI}}=0.61(4)$, with both being reduced compared to the theoretical values due to present dephasing effects. The advances presented in this work hold important promises for the implementation of advanced schemes of quantum communication using deterministic quantum light sources. 
\end{abstract}
\hspace{2cm}
\end{@twocolumnfalse}
]

\begin{figure*}[!t]
  \centering
  \includegraphics[width=\textwidth]{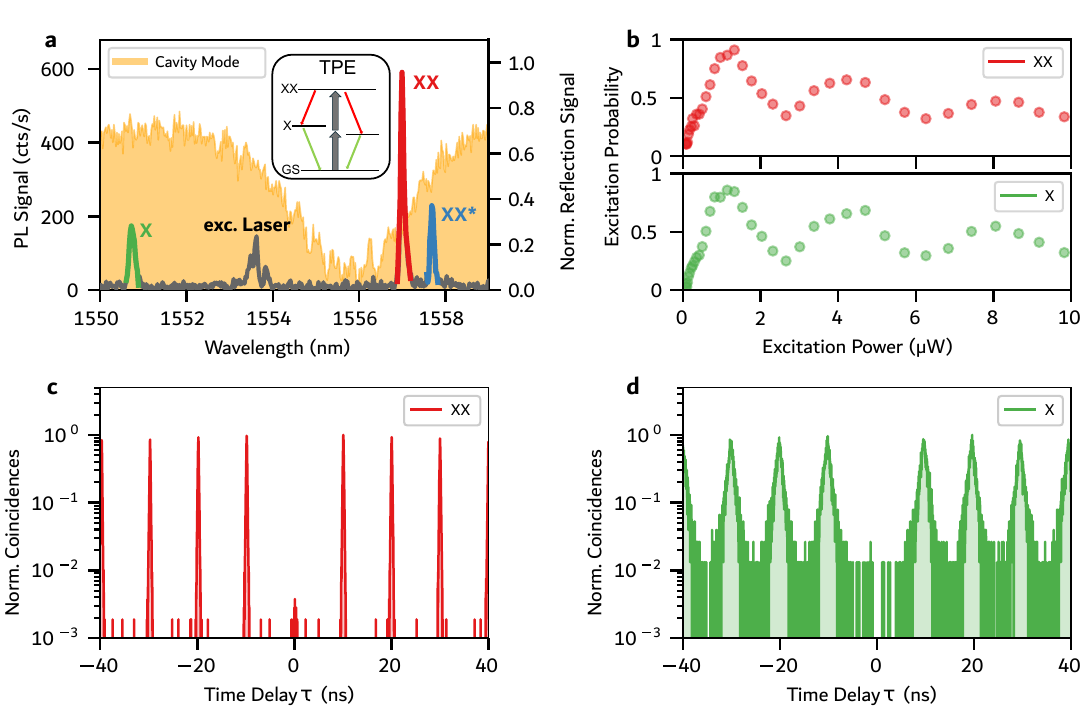}
\caption{ (a) Spectrum of the QD-device under continuous wave above-band excitation (top) and pulsed TPE (bottom). In addition to the X state at \SI{1550.4}{nm} (green) and the XX state at \SI{1557.0}{nm} (red), a third suspected charged XX* state at \SI{1557.7}{nm} (blue) is observed. The excitation laser wavelength is tuned to \SI{1553.6}{nm}, which corresponds to half of the XX-X transition energy. The cavity's reflection signal using a broadband source is shown in orange. (b) Rabi oscillations of the XX state (top) and the X state (bottom) showing the resonant excitation of the QD system. (c) and (d) second-order autocorrelation measurement of the XX and X photons (red and green respectively), showing the single-photon-nature of the presented source as integrated $g^{(2)}(0)$-values are shown.}
  \label{fig:spec_overview}
\end{figure*}

\begin{figure*}[t]
  \centering
  \includegraphics[width=0.5\textwidth]{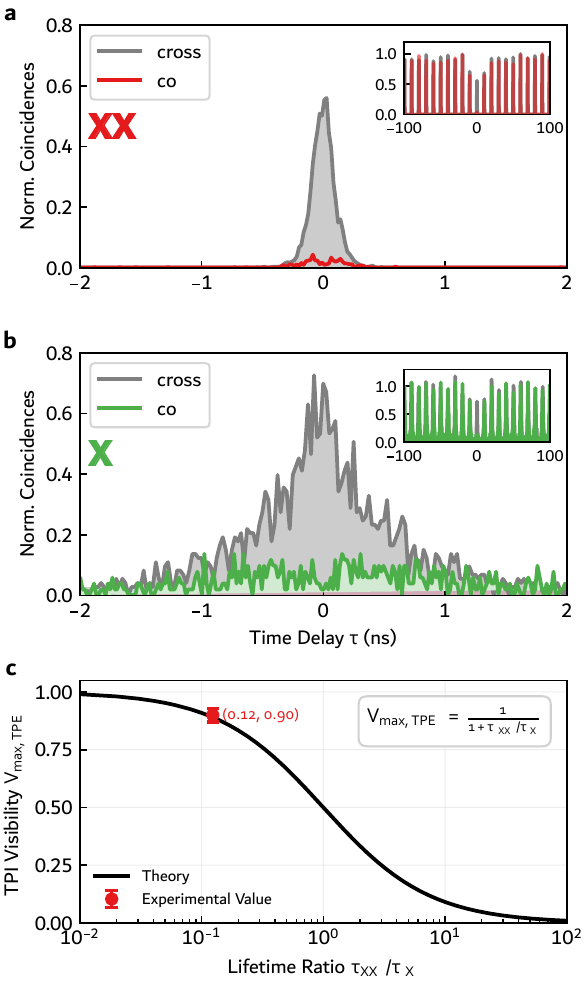}
    \caption{HOM-type TPI visibility measurement of XX (a) and X photons (b). We observe a record-high TPI visibility of the XX photons of $V_{XX} = 90(3)\%$. The X photon's visibility is impaired due to dephasing effects ($V_X = 61(4)\%$). (c) Dependence of TPI visibility on the radiative lifetime ratio. The solid black curve represents the theoretical maximum visibility $V_{\mathrm{max, TPE}}$ as a function of the ratio between biexciton ($T_1^{XX}$) and exciton ($T_1^X$) lifetimes. The experimental result (red circle) shows excellent agreement with the theoretical limit.
    }
    \label{fig:HOM_TPE}
\end{figure*}

\section{Introduction}
\label{sec:intro}
Significant advances have been made in the broad field of quantum information technologies in recent decades. This field encompasses various applications, including quantum computing, sensing, metrology and cryptography \cite{couteau_applications_2023,couteau_applications_2023-1}. When implemented with photonic hardware, many advanced protocols depend on the robust and on-demand generation of single, indistinguishable photons. While various material systems can be used for this task \cite{aharonovich_solid-state_2016, chakraborty_advances_2019}, semiconductor quantum dots (QDs)~ \cite{vajner_quantum_2022,heindel_quantum_2023} are among the most promising candidates, in terms of the achieved single-photon flux, extraction efficiency, and indistinguishability. Readily serving as close to ideal single-photon sources in the near infrared (NIR) spectral range \cite{ding_high-efficiency_2025}, the transfer of related experiments and applications to telecom wavelengths (around \SI{1550}{nm}) has attracted increasing interest, due to the compatibility with existing optical fiber infrastructures. The emission of QD-devices emitting around \SI{780}{nm} and \SI{930}{nm} is typically based on the GaAs/AlGaAs~\cite{schweickert_-demand_2018,hanschke_quantum_2018,scholl_resonance_2019,zhai_quantum_2022} and InGaAs/GaAs-~\cite{wang_near-transform-limited_2016,tomm_bright_2021} material system. By using InAs/GaAs QDs with metamorphic buffer structures~\cite{paul_single-photon_2017}, the alternative material system InAs/InP for QD growth~\cite{muller_quantum_2018,holewa_bright_2022}, also in combination with InGaAsP~\cite{phillips_purcell-enhanced_2024} and GaP barrier layers~\cite{wakileh_approaching_2025}, or InAs/InAlGaAs/InP QDs~\cite{benyoucef_telecom-wavelength_2013,kaupp_purcellenhanced_2023}, the QD emission can be shifted to the telecom C-Band. \\
These concepts have been studied in multiple works under various excitation schemes, with measured TPI visibilities ranging from 14\% to 35\% \cite{nawrath_resonance_2021, nawrath_bright_2023, holewa_high-throughput_2024, vajner_-demand_2024}, only reaching values >66\% if narrowband optical filters are applied for spectral post-selection~\cite{joos_coherently_2024}.
In combination with circular Bragg grating (CBG) resonators~\cite{yao_design_2018,liu_solid-state_2019,wang_-demand_2019,rickert_optimized_2019,barbiero_design_2022}, TPI visibilities at C-band wavelengths of up to 72\% were achieved under quasi-resonant p-shell excitation \cite{kim_two-photon_2025} and 90\% for LA-phonon excitation~\cite{hauser_deterministic_2026}. Due to the cavity interaction and faster radiative decay, the indistinguishability of Purcell enhanced emitters is less impaired by dephasing effects compared to emitters without photonic engineering~\cite{rickert_high_2025}. Moreover, the degree of indistinguishability is largely affected by the chosen excitation scheme. Strictly resonant excitation enables typically highest coherence and TPI visibilities \cite{kuhlmann_charge_2013,huber_optimal_2015}, but requires demanding spectral- or polarization-filtering, also impairing the achievable multiphoton suppression due to laser leakage and re-excitation events of emitters with fast radiative decay rates~\cite{fischer_pulsed_2018}. Two-photon resonant excitation (TPE) of the biexciton-exciton (XX-X) radiative cascade, on the other hand, suppresses re-excitation, while minimizing the requirements for spectral filtering, resulting in the highest multiphoton suppression reported to date~\cite{schweickert_-demand_2018}. But at the same time, the XX-X radiative cascade intrinsically limits the photon-indistinguishability via the lifetime ratio of the XX and X transitions~\cite{simon_creating_2005,scholl_crux_2020}. This trade-off can be resolved in two ways: (1) The XX-photons indistinguishability can be enhanced by selectively Purcell-enhancing the XX-transition relative to the X-transition~\cite{huber_measurement_2013}. (2) The X-photons indistinguishability can be enhanced using stimulated TPE (stimTPE) \cite{yan_double-pulse_2022, sbresny_stimulated_2022, wei_tailoring_2022}, where a second, timed laser pulse is used to stimulate the XX-X transition, thus eliminating the intrinsic timing-jitter of the cascaded decay.
\\
In this work, we report on TPE experiments using a cavity-enhanced telecom C-Band QD device with ultra-low XX emitter decay time and a X to XX lifetime ratio of $>$8. The asymmetric Purcell-enhancement results in high TPI visibilities approaching the theoretical limit and clearly exceeding the performance of radiative cascades without photonic engineering. Furthermore, we implement for the first time the stimTPE excitation scheme in the telecom C-band and confirm an increase in the indistinguishability of X photons, as well as an increased photon flux in the stimulated polarization state.

\section{Results and Discussion}
\label{sec:results}

\begin{table*}[!t]
\centering
\setlength{\tabcolsep}{10pt}
\begin{tabular}{@{} l l l l l @{}}
\toprule
\textbf{QD Type} & \textbf{Device} & \textbf{ES} & \textbf{Visibility} & \textbf{Ref} \\
\midrule
InAs/InAlGaAs & CBG & LA Phonon & $V_{\mathrm{raw}} = 92\%$ & \cite{hauser_deterministic_2026} \\
\textbf{InAs/InAlGaAs} & \textbf{CBG}  & \textbf{TPE} & $V_{\mathrm{raw}} = 90(3)\%$ & \textbf{[Here]} \\
InAs/InAlGaAs          & CBG           & QR           & $V_{\mathrm{raw}} = 72\%$ & \cite{kim_two-photon_2025} \\
InAs/GaAs-MB               & CBG           & LA Phonon           & $V_\mathrm{raw} = 35\%$ ($66\%)^{**}$ & \cite{joos_coherently_2024} \\
\textbf{InAs/InP}      & \textbf{Mesa} & \textbf{TPE} & $V_{\mathrm{raw}} = 29\%$ & \cite{vajner_-demand_2024} \\
\bottomrule
\end{tabular}
\caption{Reported (uncorrected) TPI visibilities $V_\mathrm{raw}$ under pulsed excitation for various C-band emitting QDs and unfiltered zero-phonon-line, but filtered phonon side band. ES: excitation scheme. MB: metamorphic buffer. $^{**}$: under strong narrow-band filtering, discarding emission from the zero-phonon-line.}
\label{tab:qd_visibility}
\end{table*}
The investigated QD-device consists of a CBG cavity containing a single InAs/$\mathrm{In_{0.53}Al_{0.23}Ga_{0.24}As}$/InP QD, grown by molecular-beam epitaxy. Details on QD growth and cavity fabrication can be found in Refs. \cite{kaupp_purcellenhanced_2023,kim_two-photon_2025}. 
Fig.~\ref{fig:spec_overview}(a) displays the QD emission spectrum under TPE (gray) as well as the cavity-mode reflection spectrum of the QD-CBG device under study. A detailed description of the experimental procedures is provided in the Methods section. The main spectral features comprise the neutral X (green) and neutral XX state (red) as well as a charged biexcitonic transition ($\mathrm{XX^*}$) (blue), where the state-assignment was based on polarization and excitation-power-dependent measurements. The cascaded nature of the XX- and X-emission was additionally confirmed via second-order cross-correlation measurements (see Supplementary Information (SI)). The coherent nature of the excitation process is demonstrated in Fig.~\ref{fig:spec_overview}(b), showing Rabi rotations of the XX and X states in the excitation-power domain up to 6$\pi$. The significant damping with increasing pulse area can be described by phonon pure-dephasing~\cite{hanschke_experimental_2025}. In the following, all measurements were recorded under $\pi$-pulse condition, if not specified otherwise. Next, the degree of multiphoton suppression is quantified by measuring the second-order autocorrelation of XX- and X-photons in a Hanbury Brown-Twiss (HBT) type setup. The corresponding histograms are depicted in Figure \ref{fig:spec_overview}(c) and (d) for the XX- and X-photons, respectively. The strongly suppressed coincidences at zero time delay $\mathrm{\tau = 0}$ clearly confirm on-demand preparation of single-photons from the QD device. To quantify the $\mathrm{g^{(2)}(0)}$ value, the histogram is fitted by a sum of double-sided exponential functions to determine the center of the integration intervals and account for blinking effects of the emitter (see SI). Then, the $g^{(2)}(0)$-value is determined by comparing the integrated events in the time-window $\pm1/(2\cdot f_\mathrm{rep}=100\,\mathrm{MHz})$ to the averaged integrated value of up to 10 neighboring coincidence peaks. 
\begin{figure*}[!t]
    \centering
    \includegraphics[width=\textwidth]{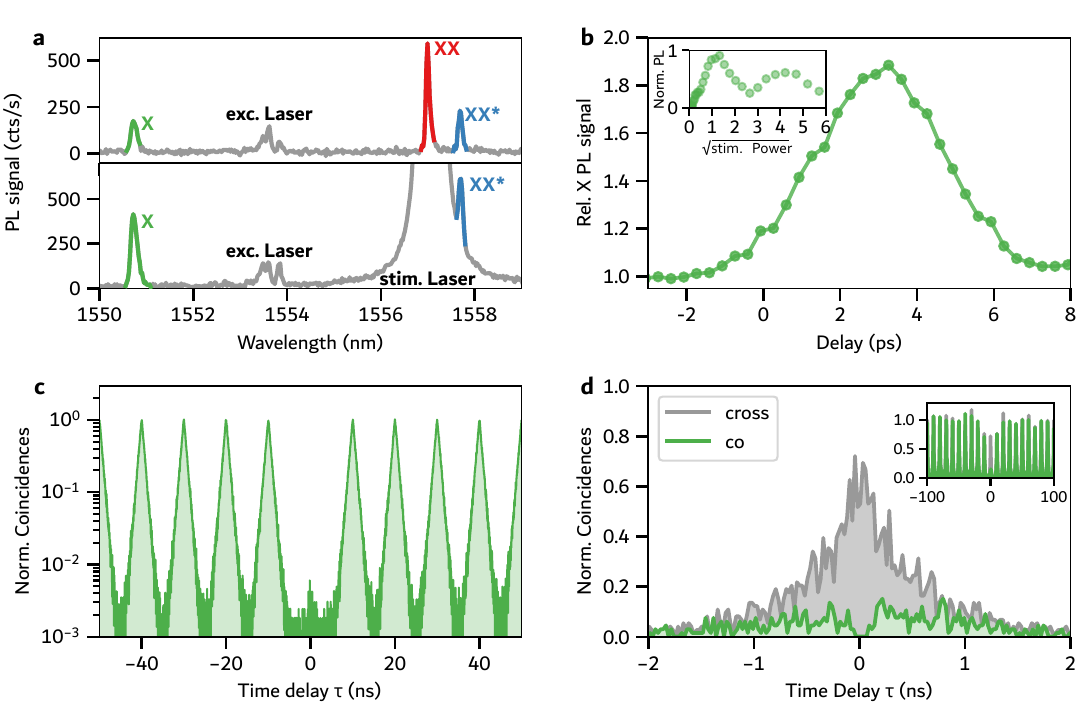}
    \caption{(a) Spectrum of the presented source under TPE (top) and stimTPE  (bottom). The stimulation laser is energetically tuned to the XX-X transition, which effectively doubles the PL signal in the stimulated polarization state. (b) Relative PL signal of the X photons as a function of delay between the excitation and the stimulation laser. The optimal stimulation window is observed at $\approx$ \SI{3}{ps}. The inset shows Rabi oscillations on the X state as a function of stimulation pulse power showing the coherent nature of the stimulation. (c) Second-order autocorrelation measurement of the X photons under stimTPE. The amount of multi-photon events is not increased in comparison to the TPE. (d) HOM-type TPI visibility measurement of the X photons under stimTPE. The TPI visibility is increased by $8(7)\%$ with respect to the TPE case.} 
    \label{fig:stim}
\end{figure*}
This yields:
 \begin{equation*}
     g_\mathrm{XX}^{(2)}(0) = 0.007(1), \ g^{(2)}_\mathrm{X}(0) = 0.006(1)
 \end{equation*}
for the XX and the X respectively. The autocorrelation histograms also reflect the degree of spectral overlap of the respective states with the cavity mode. Fitting mono-exponential decay functions to the correlation peaks in  Fig.~\ref{fig:spec_overview}(c) and (d) yields the following decay constants: 
\begin{equation*}
  T_1^\mathrm{XX} = \SI{67.4(2)}{ps}, \ T_1^\mathrm{X} =  \SI{544(2)}{ps}.
\end{equation*}
Considering the finite instrument function (IRF) of 57.9(1)\,ps (full width of half maximum) of the superconducting nanowire detector system, this yields a deconvoluted XX lifetime of 64(1)\,ps. 
While the XX-state is spectrally well aligned to the cavity mode and thus shows the lowest reported decay time for QDs in the telecom C-Band, the X-state is spectrally detuned from the cavity mode and thus shows a significantly longer decay time. However, we note that the X is still enhanced by the cavity, since bulk X states typically show decay times of about 1.2\,ns for this sample~\cite{kaupp_purcellenhanced_2023,kim_two-photon_2025}.\\
The maximum achievable TPI visibility for both XX and X photons under TPE is determined by their lifetime ratio \cite{simon_creating_2005,scholl_crux_2020} 
\begin{equation}
    V_\mathrm{max, TPE} = \frac{1}{1 + T_1^\mathrm{XX} / T_1^\mathrm{X}}.
    \label{V_theo}
\end{equation}
For the QD-device investigated here, featuring a life-time ratio of $T_1^\mathrm{X}/T_1^\mathrm{XX} >8$ between the rapidly decaying XX-state and the significantly slower X-state, a maximum TPI visibility of $V_\mathrm{max,TPE}=$ 89.47(7)\% is theoretically possible. To explore the TPI visibilities of the XX- and X-photons we conducted Hong-Ou-Mandel (HOM) type experiments. We applied a moderate spectral filtering of 70\,$\mu$eV (0.14 nm), to filter out the QDs' phonon sidebands, but not discarding emission from the zero phonon line, i.e. leaving it unfiltered. The corresponding histograms in co- and cross-polarized configuration are shown in Figure \ref{fig:HOM_TPE}(a) and (b) for the XX and X state, respectively. Evaluating the visibility according to $V = 1 - (A_\mathrm{co}/{A_\mathrm{cross}})$, using the integrated coincidence counts $A$ within a 1/(100\,MHz) time window of the respective HOM-histograms $g^{(2)}_\mathrm{HOM}(\tau=0)$ yields:

\begin{equation*}
    V^\mathrm{XX}_\mathrm{raw} = 0.90(3), \ V^\mathrm{X}_\mathrm{raw} = 0.61(4), 
\end{equation*}

The TPI visibility measured for the XX photons is (within the error limit) in excellent agreement with the theoretical prediction $V_\mathrm{max.TPE}$ (see Fig. \ref{fig:HOM_TPE}(c)), corresponding to the highest value observed to date under TPE at C-band wavelengths (Table~\ref{tab:qd_visibility}). The measured values for the TPI visibility are at the same level of the findings of Hauser et al.~\cite{hauser_deterministic_2026} observing a TPI visibility of 92\% under LA phonon excitation for a charged exciton line (which is therefore not intrinsically limited in TPI visibility according to Eq.~\ref{V_theo}) for a similar cavity-enhanced QD from the same material system. The measured TPI visibility of the X-state, on the other hand, is reduced by 31(4)\% compared to the theoretical possible value, which is the same for the XX- and X-state, limited by the state purity caused by the cascade according to Eq.~\ref{V_theo}. We attribute the reduced indistinguishability of the X photons to a larger impact of dephasing which is not captured by Eq.~\ref{V_theo}, with this dephasing being more severe for the X-transition due to the longer lifetime of the X- compared to the XX-state. 
Next, we apply the recently proposed stimulated TPE (stimTPE) scheme~\cite{yan_double-pulse_2022, sbresny_stimulated_2022, wei_tailoring_2022} to excite the X state, which was not previously shown for a QD emitting in the telecom C-band. To experimentally implement the stimTPE scheme for the first time in the telecom C-band, we introduce a second timed stimulation pulse matching the XX-emission energy, which has a temporal delay adjustable within a few picoseconds relative to the initial TPE pulse (see Methods section). The stimulation pulse ensures that the XX-X transition is de-excited via stimulated emission at a well-defined time, thus eliminating the temporal jitter otherwise resulting from the radiative decay of the XX, and thus the impairment of its TPI visibility, theoretically enabling a state purity of 1 for the X transition. We note that for the present XX-X cascade with its asymmetric decay time, the intrinsic state purity and thus TPI visibility is already close to 0.9, and only small improvements can be expected for the X in the experiment. Additionally, as was shown above, the X state is experimentally limited by spectral diffusion, resulting in a measured TPI visibility below the theoretical state purity by the cascade. We thus expect also an impaired TPI visibility under stimTPE. However, the stimTPE scheme is not only interesting regarding increased state purity for the X transition, but also because, the polarization of the stimulation pulse determines the polarization of the emitted photon of the XX-X transition, which limits the XX-X cascade to only one of the two decay paths. This allows us to control the polarization of the emitted X photons and effectively doubles the PL signal in the desired polarization state.

Figure~\ref{fig:stim}(a) shows the spectrum of the device under TPE (top) and stimTPE (bottom). The stimulation laser was tuned to the XX-X transition, with a delay of approximately \SI{3}\,ps relative to the excitation pulse. The PL signal is roughly doubled compared to the TPE spectrum of the X state for the aforementioned reasons. This increase in X emission intensity strongly depends on the delay between the excitation and stimulation pulses, as depicted in the measurement data in Fig.~\ref{fig:stim}(b). If the stimulation pulse arrives too early, the XX state has not yet been occupied, and thus cannot be stimulated. If it arrives too late, the XX state has already decayed spontaneously. For negative delays, i.e. the stimulation pulse arrives before the excitation pulse, the X emission intensity remains constant. The optimal temporal delay for stimulation is observed for $\delta \approx$3\,ps, resulting in a doubling of the PL signal in the respective polarization state. Further evidence for the successful coherent implementation of stimTPE is provided by the observation of Rabi oscillations in the X state emission intensity as a function of stimulation pulse power for pulse areas up to $4\pi$ (see inset Fig.~\ref{fig:stim}(b)). Second-order autocorrelation measurements on the X photons, as shown in Fig.~\ref{fig:stim}(c), confirm the strong suppression of multiphoton emission events with $g^{(2)}(0)_\mathrm{stim} = 0.007(1)$, agreeing within the error limit with the $g^{(2)}(0)$ measured under TPE without stimulation pulse. Finally, Fig.~\ref{fig:stim}(d) presents the corresponding HOM-type measurement on the X photons under stimTPE. The extracted value of the TPI visibility of $V^{X}_\mathrm{raw,stim}= 0.69(3)$ corresponds to an improvement of 8(7)\% compared to the result under TPE (cf. Fig.~\ref{fig:HOM_TPE}(b)). The observed 31(3)\% deviation from unity state purity aligns with the discrepancy in maximum achievable TPI visibility for X photons under TPE, which is caused by dephasing effects. Hence, the limiting factor of the X state purity is its susceptibility to dephasing mechanisms, due to its long radiative decay time. Near unity TPI visibility using the stimTPE scheme would be expected for devices with lower dephasing, whose impact could also be reduced by Purcell-enhancing the X-transition. To observe a strong improvement in visibility between stimTPE and TPE (independent of dephasing) requires a XX-cascade with a higher $T^\mathrm{XX}_1/T^\mathrm{X}_1$-ratio (i.e. smaller $V_\mathrm{max,TPE}$), for example for a device with very fast X-state decay and comparably slow XX-state decay.

\section{Conclusion}

We presented an on-demand single-photon source emitting highly-indistinguishable photons in the telecom C-band based on an InAs/$\mathrm{In_{0.53}Al_{0.23}Ga_{0.24}}$ QD embedded in a CBG cavity. The photonic engineering, with the XX-state overlapping with the cavity mode, results in a record-fast biexciton decay time of $T_1 = $ \SI{67.4(2)}{ps}, near-background-free single-photon emission ($g_{XX}^{(2)} = 0.007(1), g_{X}^{(2)} = 0.006(1)$), and a TPI visibility of the photons emitted by the XX-X transition under coherent TPE at $\pi$-pulse condition of $V^{XX}_\mathrm{raw} = 0.90(3)$, matching the theoretical expectation. Furthermore, we successfully implemented for the first time stimulated TPE at telecom C-band wavelengths and confirm a noticeable enhancement of the indistinguishability of X-photons, while maintaining excellent multi-photon suppression. Our work therefore paves the way for the generation of polarization-entangled photon-pairs with high indistinguishabilities directly in the Telecom C-band.

Future effort will concern the exploitation of deterministic fabrication technologies aiming at cavity designs with an inverted photonic engineering in combination with stimTPE, where the X-state is strongly Purcell-enhanced while the XX-state is detuned from the optical mode. This would enable brighter, polarization-programmable, and highly indistinguishable C-band photons at multi-GHz rates. Compared to very recent related work using QDs emitting at shorter wavelengths ($\approx$ 900\,nm) in combination with an open-cavity approach \cite{Baltisberger2025}, the results presented in our work feature immediate compatibility to both standard fiber-optical network infrastructures as well as methods for the efficient and robust direct fiber-pigtailing \cite{rickert_high-performance_2023,Rickert_FC-CBG_2025}.

\section{Methods}
\label{sec:methods}

The QD-device is operated in a closed-cylce 4\,K-cryostat embedded in a confocal spectroscopy setup. For implementing the TPE scheme, we employ a pulsed, spectrally-shaped C-band laser with a repetition rate of 100\,MHz (center wavelength: 1560\,nm, native pulse duration: $<$90\,fs). A pulse-slicer is used to reduce the spectral width of the laser pulses to \SI{0.8(1)}{nm}, corresponding to a pulse-duration of 4.4\,ps, and match the excitation energy to the two-photon absorption resonance of the XX state. The XX-state subsequently decays via the XX-X-G cascade, emitting a state which can be written as $\ket{\psi}=\frac{1}{\sqrt{2}}\left(\ket{\mathrm{H_XH_{XX}}}+e^{i\delta\tau/\hbar}\ket{\mathrm{V_{XX}V_X}}\right)$ yielding a correlated photon pair when filtering along the QD's eigenaxis in H/V polarization \cite{liu_solid-state_2019, wang_-demand_2019, muller_-demand_2014}. To suppress the excitation laser a set of three notch filters (center wavelength: 1555\,nm, FWHM:  1\,nm) is used.
In the stimTPE scheme, the initial laser beam is split into an excitation and stimulation beam using a half-wave plate (HWP) and a polarizing beam splitter (PBS). Two separate pulse shapers adjust the spectral widths and pulse energies. Additionally, the stimulation pulse power can be controlled using a variable optical attenuator (VOA), while a delay stage controls the temporal delay between the excitation and stimulation pulses. The two paths are then recombined using a fiber beam splitter.
For photon detection, a grating spectrometer with attached InGaAs-array camera can be used. Alternatively, for TPI and autocorrelation measurements, the InGaAs-array camera is bypassed to couple the desired emission line into a HOM-setup. This setup is connected to a single-photon detection system based on superconducting nanowires and time-tagging electronics (overall temporal instrumental response function (IRF) with  57.9(1)\,ps full width half maximum), for time-resolved and (auto)correlation measurements.

\section{Acknowledgments}
The authors gratefully acknowledge early contributions to the experimental methodology, software and helpful discussions by D. A. Vajner, as well as expert sample fabrication by M. Emmerling.

\section{Data and Availability}
The data presented in this work is available from the authors upon reasonable request.

\section{Funding}
The authors acknowledge financial support by the German Federal Ministry of Research, Technology and Space (BMFTR) via the project “QuSecure” (Grant No. 13N14876) within the funding program Photonic Research Germany, the BMFTR joint projects “tubLAN Q.0” (Grant No. 16KISQ087K) as well as QuNET+ICLink (Grant No. 16KIS1967 and 16KIS1975) in the context of the federal government’s research framework in IT-security “Digital. Secure. Sovereign.”. The authors further acknowledge funding by the State of Bavaria. T. H.-L. acknowledges financial support by the BMFTR via the project "Qecs" (Grant No. 13N16272). A. T. P acknowledges funding by the BMFTR via the project "Ferro35" (Grant No. 13N17641).

\section{Disclosure}
The authors declare no conflict of interest.

\bibliographystyle{ieeetr}
\bibliography{references-2}

\end{document}